# Improved Thermoelectric Properties in (1-$x$)LaCoO$_3$/($x$)La$_{0.7}$Sr$_{0.3}$CoO$_3$ Composite

Ashutosh Kumar[1,*], D. Sivaprahasam[2], Ajay D Thakur[1,**]

[1]Department of Physics Indian Institute of Technology Patna Bihta 801 106 India
[2]Center for Automotive Energy Materials ARC International IITM Research Park Chennai 600 113 India

**Abstract:** A high Seebeck coefficient ($S$), large electrical conductivity ($\sigma$), and reduced thermal conductivity ($\kappa$) are required to achieve a high figure-of-merit ($zT$) in an ideal thermoelectric (TE) system, which is challenging in a single system due to the interdependence of TE parameters. Composite approach is promising to manipulate the TE parameters. In this study, TE properties of (1−$x$)LaCoO$_3$/($x$)La$_{0.7}$Sr$_{0.3}$CoO$_3$ (0.00≤x≤0.05) composite is discussed. The structural analysis confirms individual phases in the composite, which is further supported by electron microscopy analysis. The x-ray photoelectron analysis indicates that oxygen vacancies (VO) are present in the parent LaCoO$_3$ system and increase with the addition of La$_{0.7}$Sr$_{0.3}$CoO$_3$ (LSCO) in the composite. The increase in VO raises the degenerate states of cobalt and hence improves S in the composites. Temperature variation in $S$ and $\sigma$ are consistent with the spin-state transition and shows the correlation between these two parameters. The reduction in κ and σ with the addition of ball-milled La$_{0.7}$Sr$_{0.3}$CoO$_3$ in the composite is attributed to the enhanced phonon-phonon and charge carrier scattering, respectively. A synergistic effect of enhanced $S$ and reduced $\kappa$ result in five times improvement in $zT$ of the composite compared to the parent LaCoO$_3$ system at 800 K. This approach also improves the operating temperature for LaCoO$_3$ based systems.

[*]Email : science.ashutosh@gmail.com, present address: Lukasiewicz Research Network - Krakow Institute of Technology Krakow, Poland
[**]Email: ajay.thakur@iitp.ac.in

# INTRODUCTION

Oxide thermoelectric (TE) materials are a potential choice for direct conversion of heat to electrical energy for power generation and cooling applications owing to their stability at high temperatures in an open environment [1–3]. The efficiency of these TE power generators depends on figure-of-merit ($zT$, a dimensionless quantity) and $zT$ depends on Seebeck coefficient ($S$), electrical conductivity ($\sigma$), thermal conductivity ($\kappa$), and absolute temperature ($T$) as $zT = S^2\sigma T/\kappa$, where $S^2\sigma$ is commonly known as power factor. It is often challenging to optimize these three parameters in a single material with low $\kappa$ (such as in glass), high $\sigma$ (like in metals), and high $S$ (as in semiconductors) owing to their interdependence. Also, $\kappa$ is a combination of two parts: electronic thermal conductivity ($\kappa_e$) and phonon thermal conductivity ($\kappa_{ph}$) where $\kappa_{ph}$ dominates the $\kappa$, in general, for oxide TE materials. Several methodologies are employed to enhance the $zT$ value in oxide TE systems by improving $S^2\sigma$ and reducing the $\kappa_{ph}$ [4–6].

Cobalt-based oxides ($Na_xCoO_2$ [7], $LaCoO_3$ [8,9], etc.) are capable of showing improved TE properties due to the possibility of having different charge states of cobalt (Co) accompanying with different spin-states, which tunes $S$ following Heikes formula [10]. The electronic configuration of Co in different spin-states are $Co^{3+}$: LS ($t_{2g}^6 e_g^0$), IS ($t_{2g}^5 e_g^1$), HS ($t_{2g}^4 e_g^2$) and $Co^{4+}$: LS ($t_{2g}^5 e_g^0$), IS ($t_{2g}^4 e_g^1$), HS ($t_{2g}^3 e_g^2$) where LS, IS and HS are low spin, intermediate spin, and high spin, respectively [11]. The transition between these spin-states is temperature dependent [12,13]. Also, the spin-state transition temperature depends on alkaline metals' substitution at rare earth sites in $LaCoO_3$ [14,15]. The $zT$ values in $LaCoO_3$ (LCO) can be improved by co-substitution and a composite approach [16–18]. However, in these studies, $\kappa$ values are dominated by $\kappa_{ph}$. It indicated a possibility to further improve the $zT$ values by reducing the $\kappa_{ph}$, which can be achieved in nanostructured systems.

Motivated by the dominant nature of $\kappa_{ph}$ in the $LaCoO_3$ system, as reported in our previous studies [16–18], we study the TE properties in $(1-x)LaCoO_3/(x)La_{0.7}Sr_{0.3}CoO_3$ composite. The secondary phase ($La_{0.7}Sr_{0.3}CoO_3$) in the present study differs from previous studies [17,18]. This conducting phase is also ball-milled to reduce its particle size further, enhancing the scattering of phonon to reduce the system's lattice thermal conductivity.

# EXPERIMENTAL SECTION

Polycrystalline samples of $LaCoO_3$ (LCO) and $La_{0.7}Sr_{0.3}CoO_3$ (LSCO) are synthesized using the standard solid-state method. In the synthesis process, the stoichiometric amount of $La_2O_3$, $Co_3O_4$ for LCO and $La_2O_3$, $SrCO_3$ & $Co_3O_4$ for LSCO are weighed and mixed separately using mortar-pestle in acetone as a liquid medium. The mixture of LCO and LSCO is calcined separately in a muffle furnace in the air using alumina crucible at 1473 K for 20 hours with 3K/min heating and cooling rate. Among the LCO and LSCO calcined samples, the LSCO sample is ball-milled to reduce its particle size using a planetary ball mill with 300 rpm for 24 hours. The ball (zirconia) to powder weight ratio is maintained to 10:1. The ball-milled LSCO is then added with LCO to make $(1-x)LaCoO_3/(x)La_{0.7}Sr_{0.3}CoO_3$ composite for $x$=0.00, 0.02 and 0.05, $x$ represents the weight fraction. It is noted that our previous reports show a high lattice thermal conductivity in the composite materials [17,18]. Hence, the second phase ($La_{0.7}Sr_{0.3}CoO_3$, different from previous studies) is ball-milled to improve the phonon-phonon scattering. These LCO-LSCO mixtures are then consolidated into a cylindrical pellet of 10 mm diameter (under 25 MPa). The compacted pellets were sintered at 1523 K for 12 hours in the muffle furnace in air. Crystallographic structure and phase identification are done via x-ray diffraction (XRD) pattern using the Rigaku diffractometer ($\lambda$=1.5406 Å) followed by Rietveld refinement. Field emission scanning electron microscopy (FE-SEM, Zeiss) and energy-dispersive X-ray spectroscopy (EDXS) are performed for microstructural and chemical composition studies. High-resolution transmission electron microscope (HRTEM) is used to demonstrate the presence of LCO and LSCO in the composite. Further, X-ray photoelectron spectroscopy (XPS) measurement (Thermo Fischer scientific ESCALABT M Xi+) is done at room temperature to investigate the composite's chemical composition. The sintered pellet is cut into a rectangular bar of 12 mm× 4mm × 4mm dimension to measure $\sigma$ and $S$ from the Seebsys$^{TM}$ system in a four-probe configuration in a wide temperature range from 300 K to 800 K. The thermal conductivity ($\kappa$=D$\rho$C$_v$) was measured using laser flash analysis (LFA), where D is thermal diffusivity, $\rho$ is sample density (calculated using the sample mass and its geometric volume), and C$_v$ is specific heat (calculated using Dulong-Petit law). Potential Seebeck microprobe (PSM, PANCO Gmbh, Germany) measurement is used to probe the local variation in $S$ for all the samples at room temperature [19]. The typical sample dimensions are 2 mm × 1 mm, and PSM measurements are performed with a position resolution of 25 μm.

## RESULTS AND DISCUSSION

Figure 1(a) depicts the powder x-ray diffraction (PXRD) pattern for (1−$x$)LaCoO$_3$/($x$)La$_{0.7}$Sr$_{0.3}$CoO$_3$ composite with $x$=0.00, 0.02-BM and 0.05-BM. The diffraction pattern of the composite shows the XRD peaks corresponding to LCO and LSCO only. The XRD pattern for the LaCoO$_3$ perovskite system shows a rhombohedral structure having the characteristic 2θ peaks at 32.88° (110) and 33.24° (104) and is in line with the ICDD File no. 048-0123 [20]. Figure 1(b) represents the XRD pattern for the ball-milled LSCO sample. The maximum intensity peak corresponding to (110) and (104) for LSCO is indistinguishable as the broadening due to crystallite size leads to the superposition of these two peaks. In the composite sample, the diffraction pattern due to LSCO is not visible: (i) due to its small weight fraction in the composite, (ii) the characteristic 2θ peaks for LSCO are similar to LCO. However, a shift in the 2θ position of the maximum intensity peak towards a lower 2θ angle for $x$=0.02-BM is observed. The 2θ position's shift remains unchanged for the sample with $x$=0.05-BM (inset Fig. 1a). The shift in the 2θ position is similar for all the peaks. This is understood as following: according to ICDD Files, the LaCoO$_3$ has peaks at 32.88° (110), and 33.24° (104) (048-0123), and La$_{0.7}$Sr$_{0.3}$CoO$_3$ has peaks at 32.89° (110) and 33.12° (104) (075-8570). The superposition of these peaks in the composite shift the maximum intensity peak (104) towards lower 2θ values and (110) towards higher 2θ values, and hence the two peaks come closer to each other. Further, an increase in the LSCO phase fraction in the composite does not change the 2θ position; rather, it broadens the peak and is ascribed to the nanocrystalline phase's increase in the composite. It is observed that the 2θ position for the pristine LCO sample is 33.24°, which decreases to 33.20° for $x$=0.02-BM and remains constant for $x$=0.05-BM. This suggests that both the composite phases are preserved and consistent with the previous studies in similar composites [17,18,21]. However, due to sintering at high temperatures, there might be some inter-atomic diffusion between the two components in other composite systems reported elsewhere [22]. However, in the present study, for a small LSCO phase fraction (up to 5%), there is no diffusion of LCO, and LSCO phases are seen.

The powder XRD pattern of (1−$x$)LaCoO$_3$/($x$) La$_{0.7}$Sr$_{0.3}$CoO$_3$ composite is further analyzed using the two-phase Rietveld refinement procedure Fullprof$^{TM}$ software with R-3c space group to estimate the structural information such as lattice parameters, phase fraction etc. The refinement pattern for the composite sample with $x$=0.00, $x$=0.02-BM, and $x$=0.05-BM is shown in Fig. 2(a-c). The corresponding lattice parameters, phase fraction for each phase, and refinement parameters are shown in Table I. The lattice parameters for LCO obtained from the refinement pattern is similar to that observed in previous reports [23]. It is noted that the individual phases of LCO and LSCO are refined first using their powder XRD pattern, and the lattice parameters obtained from these refinements have been used to simulate the powder x-ray diffraction pattern of the composite samples. As observed from the XRD pattern, the lattice parameters are kept constant during the two-phase Rietveld refinement of the composite samples. The goodness of fit for all the refinement is well within the acceptable limit. The phase fraction obtained from the refinement is in line with the nominal composition of the composite.

Field emission scanning electron microscope (FESEM) images for $LaCoO_3$, ball-milled $La_{0.7}Sr_{0.3}CoO_3$ and $(1-x)LaCoO_3/(x) La_{0.7}Sr_{0.3}CoO_3$ powders for $x$=0.05-BM are shown in Fig. 3(a-d). Fig. 3(a) shows the large particle size of LCO due to the high calcination temperature. The corresponding particle size is in the range of 0.25-3 µm, as shown in the particle size distribution in the inset of Fig. 3(a). Fig. 3(b) shows the FESEM images of the ball-milled $La_{0.7}Sr_{0.3}CoO_3$ sample. It is observed that the particle size for $La_{0.7}Sr_{0.3}CoO_3$ lies in the range of ~20-500 nm. The distribution curve corresponding to ball-milled LSCO is shown in the inset of Fig. 3(b). Fig. 3(c) shows the FESEM image of the composite with $x$=0.05-BM. Since the weight fraction of LSCO in the composite is small (~5%), it is difficult to find a large number of small particle size in the FESEM image. However, a few particles with an average diameter of ~ 200 nm (Fig. 3(d)) are observed along with LCO phase. $LaCoO_3$ based materials display an extensive distribution of particle sizes, in general, and hence it is difficult to conclude that this small particle is of LSCO and not LCO. We performed EDXS measurements at two different regions on the zoom-in sample (Fig. 3(d)). Fig. 3(e) shows the EDXS spectra of the $LaCoO_3$ particle (region 1), and as can be seen from the spectra, peaks corresponding to La, Co, and O are observed. Also, the corresponding atomic fraction of each element is shown and is similar to the $ABO_3$ structure. Fig. 3(f) shows the EDXS spectra of the small particle (~194 nm, Fig. 3(d)), and it is noted that along with La, Co, and O, we also observed a tiny peak due Sr, which confirms that the small particle is of LSCO. This confirms the existence of LCO and LSCO phases in the composite. The elemental composition obtained for $(1-x)LaCoO_3/(x) La_{0.7}Sr_{0.3}CoO_3$ composite from EDXS measurement is shown in Table II. Further, the density measurements for all the pellets after sintering have been performed. The experimental density, theoretical density, and relative density for all the composite samples are shown in Table III. Theoretical densities for the composite samples are calculated following the rule of mixtures. The experimental densities are calculated using the mass of the sintered pellets and their geometric volume. It is found that the relative density for these samples lies between 86-88%. This agrees with the relative density obtained in the literature for cobaltate system using the conventional sintering process [24].

In order to further investigate the existence of LCO and LSCO in the composite, we performed high-resolution transmission electron microscope (HRTEM) measurement and is shown in Fig. 4. Fig. 4(a) depicts the TEM image for $(1-x)LaCoO_3/(x)La_{0.7}Sr_{0.3}CoO_3$ composite with $x=0.02$-BM. The large particle size of LCO is observed along with few small particle size of LSCO. Further analysis of the HRTEM images using ImageJ software depicts two lattice spacing (0.272 nm and 0.192 nm) as shown in Fig. 4(b). It is worth noting that, the lattice spacing 0.272 nm corresponds to the 2θ peak of LCO at 32.88° (100) and that of 0.192 nm is in agreement with the 2θ peak of LSCO at 47.25° (024). This confirms the presence of two phases in the composite, which is further supported by elemental mapping. TEM image for $(1-x)LaCoO_3/(x)La_{0.7}Sr_{0.3}CoO_3$ composite shown in Fig. 4(c) is further analyzed using EDXS. The elemental mapping corresponding to the region marked in Fig. 4(c) is shown in Fig. 4(d-g). The presence of Sr map confirms that the small particle size corresponds to LSCO in the composite.

In order to investigate the chemical composition of the samples, wide and short-range XPS measurements are performed at 300 K. Fig. 5(a-f) represents the XPS peak for the sample with $x = 0.00$ and 0.05-BM samples. Fig. 5(a) represents a wide scan range for both samples in which all the individual peaks of the corresponding elements are observed. In Fig. 5(b), a small peak in the Sr 3d spectra for the sample with $x = 0.05$-BM is observed, absent in the parent sample ($x=0.00$). This may be attributed to the small addition of $La_{0.7}Sr_{0.3}CoO_3$ to the LCO matrix. Further, the short-range scan of Co-2p spectra for both the samples is shown in Fig. 5(c-d). The XPS peaks for Co-$2p_{3/2}$ and Co-$2p_{1/2}$ are observed at 780.6 eV and 795.8 eV, respectively, for both the samples. The deconvolution of the Co-$2p_{3/2}$ and Co-$2p_{1/2}$ peaks shows the contribution of $Co^{3+}$ and $Co^{2+}$ [25]. It is noted that in $LaCoO_3$ sample, the charge state of cobalt should be $Co^{3+}$, in general, and in $La_{0.7}Sr_{0.3}CoO_3$, the fractional addition of Sr at La site leads to change the same proportion of $Co^{3+}$ to $Co^{4+}$. The binding energy corresponding to $Co^{3+}$ and $Co^{4+}$ are very close [26] and are shown in a single peak due to $Co^{3+}$ in Fig. 5(c-d). In pristine $LaCoO_3$, along with $Co^{3+}$, there is a peak due to $Co^{2+}$, which indicates that the pristine $LaCoO_3$ possesses oxygen deficiency [27].

Further, the intensity corresponding to $Co^{2+}$ increases along with $Co^{3+}/Co^{4+}$ with an increase in the LSCO phase fraction in the composite, suggesting a further increase in the oxygen vacancy in the system. $Sr^{2+}$ substitution at $La^{3+}$ site in $LaCoO_3$ also leads to the oxygen vacancies [28]. Hence, the addition of LSCO in the composite shows a further increase in oxygen vacancies.

The individual peak of O-1s is de-convoluted into three distinct peaks having binding energies of 529.2 eV, 531.3 eV, and 532.6 eV, respectively for $x = 0.00$, as shown in Fig. 5(e). The oxygen peak at 529.2 eV is attributed to oxygen atoms at the regular lattice sites (LO). The oxygen 1s peak having binding energy 531.3 eV corresponds to the oxygen-deficient (VO) regions, and the oxygen peak having binding energy higher than VO is attributed to the presence of interstitial oxygen (IO) atoms or non-stoichiometric oxygen [29]. The presence of the VO peak in the O-1s spectra for LCO confirms oxygen vacancy in the pristine sample and supports the XPS peak due to $Co^{2+}$ in LCO. Further, the oxygen vacancies are found to increase in the composite and are observed from Fig. 5(e-f), i.e., the intensity of the regular oxygen sites O-1s peak decreases and O-1s peak for oxygen vacancies increases for $x = 0.05$-BM composite sample. This suggests increasing the $Co^{2+}$ charge states in the composite, as observed in Fig. 5(c). In the present composite system, the different charge states of Co do exist and are confirmed from the XPS measurements.

Figure 6 shows the Seebeck coefficient ($S$) temperature variation for the composite with $x=0.00$, $x=0.02$-BM, and $x=0.05$-BM from 300-800 K. At 300 K, the value of S is positive for $LaCoO_3$, indicating the p-type conduction in the system dominated by the presence of a hole. Hebert et al. showed that the sign of the Seebeck coefficient in $LaCoO_3$ could be tuned depending upon the extent of hole and electron doping [30]. The S value for $LaCoO_3$ at 300 K is 236 µV/K. This is lower than several reported studies in the pure LCO system at 300 K [8,9]. This decrease in $S$ for LCO may be due to oxygen vacancies in the LCO system itself, as confirmed from XPS measurement, and is similar to previous studies [31]. Further, $S$ increases from 236 µV/K (x=0.00) to 245 µV/K (x=0.02-BM) to 256 µV/K (x=0.05-BM) at 300 K. This increase in S with the addition of LSCO ball-milled sample may be ascribed to the increase in the oxygen vacancies (VO). This increase in VO leads to improve the degenerate states in cobalt and hence improves the Seebeck coefficient.

The Seebeck coefficient in $LaCoO_3$ originates from the strong electronic correlation and large configurational entropy in cobaltates. Several spin-states and orbital degeneracies in cobaltate drive the configurational entropy, which improves the Seebeck coefficient. Further, the *S* for $LaCoO_3$ system first increases with the increase in the temperature up to 340 K and decreases rapidly with further increase in the temperature as reported elsewhere [31]. Slope change in the *S* curve's temperature dependence is observed at 340 K and 640 K and is attributed to the system's spin-state transition [32]. The first slope change may be due to the transition from the IS state to a mixed state of IS and HS state, and the transition at 640 K is due to the change in the mixed state of IS and HS to HS state [32].

In the composite samples, *S* also increases up to 360 K and then decreases slowly compared to the parent LCO sample with a further increase in the temperature. We observed a change in the spin-state transition temperature in the composite consistent with the previous reports, which suggest that the transition temperature depends on the substitution of the alkaline element at the rare-earth site [33]. In addition to this, we also observed a slope change in *S* dependence on temperature in composite samples. This is due to the temperature variation of the spin-state transition in the substituted and composite samples. We also observed that with the addition of LSCO, the temperature at which the slope changes are observed is shifted to a higher temperature, which shows that the intermediate spin (IS) state is active up to higher temperatures in the composite samples. This may be understood as following: As observed from the XPS measurement, the addition of LSCO leads to a decrease in the oxygen concentration and hence increases the concentration of different charge states of cobalt. This leads to an increase in the charge states' degeneracy, and hence the energy required to change the spin states is more, and therefore the transition temperature increases. The value of *S* is 138 µV/K at 800 K for LSCO added sample (x=0.05-BM), which is around four times higher than the LCO sample and is the maximum S values compared to other reports on co-substituted and composite based LCO systems [34,35]. The Seebeck coefficient for Sr and Mn substituted LCO is ∼ 30 µV/K at 800 K and is similar for other single or co-substituted samples [16]. However, in composite samples, the Seebeck coefficient values are ∼ 60 µV/K at 800 K [18] and are attributed to a decrease in electrical conductivity.

The Seebeck coefficient obtained using a four-probe configuration provides the average contribution of the sample at each temperature. However, the spatially resolved distribution of the *S* is measured using a potential Seebeck microprobe. Local variation of *S* at 300 K for the composite samples with $x=0.00$, $x=0.02$-BM, and $x=0.05$-BM are shown in Fig. 7(a-c). The distribution of the $S$ across the sample surface is also shown from the histogram curve in Fig. 7(d-f). For $x=0.00$, we observe a small variation in $S$, as shown from the histogram curve in Fig. 7(d), which suggests the homogeneous nature of the LCO sample. The histogram for the composite samples is broader than LCO, suggesting that both phases in the composite results in a broad distribution of $S$, as shown in Fig. 7(e-f). We did not observe smaller $S$ values, which correspond to LSCO, due to the difference in particle size (~20-500 nm) and PSM resolution (25µm). The average value of S obtained from PSM for $(1-x)$LaCoO$_3$/$(x)$ La$_{0.7}$Sr$_{0.3}$CoO$_3$ with $x=0.00$, 0.02-BM and 0.05-BM are 231.2 µV/K, 241.4 µV/K and 249.3 µV/K respectively. These $S$ values observed for all the samples are consistent with the Seebeck coefficient measurement at 300 K. Also, the results obtained from PSM are consistent with the previous studies in similar composite systems [17,18].

Figure 8 shows the variation of electrical conductivity ($\sigma$) for the composite samples with $x=0.00$, 0.02-BM, and 0.05-BM from 300-800 K measured using the standard DC four-probe technique. The σ of composite decreases with the increase in LSCO phase fraction at 300 K. The decrease in σ may be due to an increase in charge carrier scattering due to an increase in the nanocrystalline phase [36,37]. The $\sigma$ at 300 K for LCO is 0.15 S/cm, and it reduces to 0.09 S/cm for $x=0.02$-BM to 0.05 for $x=0.05$-BM. Further, $\sigma$ is found to increase with the increase in temperature for all the samples showing the semiconducting behavior. The $\sigma$ decreases by order of magnitude at 800 K with ball-milled LSCO in the composite. In the present case, the $\sigma$ at 800 K is ~632 S/cm for LaCoO$_3$, and it reduces to 91 S/cm for $x=0.02$-BM to 52 S/cm for $x=0.05$-BM. It is noted that σ for $(1-x)$LaCoO$_3$/$(x)$La$_{0.95}$Sr$_{0.05}$CoO$_3$ composite at 800 K is similar to parent LaCoO$_3$ for $x=0.05$ when the particle size of both the phases are similar and decreases with higher $x$ [18]. However, despite using the more conducting second phase (La$_{0.7}$Sr$_{0.3}$CoO$_3$) in the present study, $\sigma$ decreases significantly and may be attributed to the reduced particle size, enhancing the charge carrier scattering. This decrease in the σ values for the composite samples are analogous to the increase in S values at the same temperature. We show the Arrhenius plot (ln $\sigma$ vs. 1000/$T$) for the composite samples in the inset of Fig. 8. Slope changes are observed in the Arrhenius plot and may be correlated to the spin-state transition in the cobalt [38]. There are three different regions observed in the linear fitted curve for the composites with $x=0.00$, 0.02-BM, and 0.05-BM and analogous to the slope change in the temperature-dependent Seebeck coefficient. This also shows a strong correlation between $\sigma$ and $S$.

When the electrical transport is determined by one type of carrier (in the present study, the majority of carriers are holes), in that case, the total thermal conductivity $\kappa$ is a combination of two components: electronic thermal conductivity ($\kappa_e$) and phonon thermal conductivity ($\kappa_{ph}$). The $\kappa_e$ can be calculated by Wiedemann-Franz law: $\kappa_e = L\sigma T$, where $L$ is the Lorenz number and is calculated using the $L = 1.5 + \exp[-|S|/116]$ [39]. Further, $\kappa_{ph}$ has been extracted by subtracting $\kappa_e$ from $\kappa$. The $\kappa$, $\kappa_e$, and $\kappa_{ph}$ as a function of temperature for $(1-x)$LaCoO$_3$/$(x)$ La$_{0.7}$Sr$_{0.3}$CoO$_3$ composite sample are shown in Fig. 9. At 300 K, $\kappa$ reduces slightly with the increase in ball-milled LSCO phase fraction in the composite. However, with the increase in temperature, we observed a sharp decrease in $\kappa$ in the ball-milled sample. $\kappa_e$ increases with temperature for LCO due to an increase in the corresponding $\sigma$. Further, $\kappa_{ph}$ increases with the increase in temperature up to ~ 550 K, and then remains constant and decreases. This decrease may be attributed to the decrease in the mean free path at higher temperatures. For the composite sample, we obtained a decrease in $\kappa_{ph}$, which may be attributed to the enhanced phonon scattering due to the small particle size LSCO in the composite [40–42]. Hence, a decrease in $\kappa$ for the composite is attributed to the decrease in both components, i.e., $\kappa_e$ due to decreased $\sigma$, $\kappa_{ph}$ due to enhanced phonon scattering. The $\kappa$ at 800 K for the LaCoO$_3$ sample is ~ 2.9 W/m-K, and it lowers to ~ 1.48 W/m-K at the same temperature for $x$=0.02-BM and 1.40 W/m-K for $x$=0.05-BM. This suggests that the $\kappa$ of LCO can be reduced almost to its half values by adding a small phase fraction of the nanocrystalline LSCO phase. This significant decrease in $\kappa$ with the nanocrystalline LSCO phase may be attributed to interface thermal resistance's prominent effect with reduced particle size [43]. A comparison of $\kappa$ observed in a similar cobaltate system is shown in Table IV.

The TE parameters $S$, $\sigma$, and $\kappa$ are used to estimate the composite samples' figure-of-merit ($zT$). The $zT$ as a function of temperature for $(1-x)$LaCoO$_3$/$(x)$La$_{0.7}$Sr$_{0.3}$CoO$_3$ composite is shown in Fig. 10. The $zT$ value of the LSCO added sample is interestingly found to increase at a higher temperature compared to the LCO system. This may be attributed to the synergistic effect of an increase in $S$ and decreased $\kappa$ at a higher temperature than the LCO system [16]. We observed a maximum $zT$ of ~ 0.064 at 800 K for $(1-x)$LaCoO$_3$/$(x)$ La$_{0.7}$Sr$_{0.3}$CoO$_3$ with $x$=0.02-BM sample. This is higher than several studies in LCO based systems at higher temperatures, as shown in Table

IV. This indicates that the simultaneous effect of improving $S$ and reducing $\kappa$ is promising for improving the oxide system's TE performance at higher temperatures, which is essential to elicit higher efficiency and provides a pathway for future studies in composite thermoelectric.

# CONCLUSION

Thermoelectric properties of $(1-x)$LaCoO$_3$/$(x)$La$_{0.7}$Sr$_{0.3}$CoO$_3$ composite, synthesized using standard solid-state route, is investigated in the present study. The x-ray diffraction analysis confirms the existence of both the phases in the composite, which is further supported by scanning electron microscopy and high-resolution transmission electron analysis. The x-ray photoelectron spectroscopy analysis reveals the presence of oxygen vacancies (VO) in the parent LaCoO$_3$ system which increases in the composite. Further increase in $S$ at 300 K with the LSCO phase's addition is observed and attributed to the increase in cobalt degenerate state with increasing oxygen vacancies. At higher temperatures, $S$ in the composite is significantly improved. This may be ascribed to the following: (a) decrease in $\sigma$, due to enhanced charge carrier scattering (b) increase in spin-state transition temperature. The variation in $\sigma$ with temperature for all the samples shows semiconducting nature. The slope changes in the Arrhenius plot (ln $\sigma$ vs. 1/$T$) are analogous to the composite spin-state transition. Further, a significant reduction in $\kappa$ with increasing LSCO phase fraction in the composite is observed. This shows that the nanostructured LSCO system also acts as a phonon scattering center that enhances the phonon-phonon scattering and reduces $\kappa_{ph}$. The simultaneous enhancement of $S$ and reduction in $\kappa$ results in an improved $zT$. A maximum $zT$ of $0.064\pm0.007$ at 800 K for $(1-x)$LaCoO$_3$/$(x)$ La$_{0.7}$Sr$_{0.3}$CoO$_3$ with $x=0.02$-BM is observed, which is five times higher than the LaCoO$_3$ system at 800 K. This study shows that zT in oxide composite materials can be improved by reducing the thermal conductivity. The reduction in $\kappa$ can be further optimized by considering the interface thermal resistance and the Kapitza radius [43][44] with a minimal reduction in $\sigma$, which can be beneficial in further improving figure-of-merit.

# ACKNOWLEDGMENTS

A. Kumar and A. D. Thakur would like to acknowledge the ministry of human resource and development (MHRD) India for financial support.

# DECLARATION OF COMPETING INTEREST

The authors have no conflict of interest to disclose.

# References

[1]     T.M. Tritt, M.A. Subramanian, Thermoelectric Materials, Phenomena, and Applications: A Bird's Eye View, MRS Bull. 31 (2006) 188–198. https://doi.org/10.1557/mrs2006.44.

[2]     C. Wood, Materials for thermoelectric energy conversion, Rep. Prog. Phys. 51 (1988) 459–539. https://doi.org/10.16309/j.cnki.issn.1007-1776.2003.03.004.

[3]     J.W. Fergus, Oxide materials for high temperature thermoelectric energy conversion, J. Eur. Ceram. Soc. 32 (2012) 525–540. https://doi.org/10.1016/j.jeurceramsoc.2011.10.007.

[4]     H. Ohta, K. Sugiura, K. Koumoto, Recent progress in oxide thermoelectric materials: P-type Ca 3Co4O9 and n-Type SrTiO3-, Inorg. Chem. 47 (2008) 8429–8436. https://doi.org/10.1021/ic800644x.

[5]     M.G. Kanatzidis, Nanostructured thermoelectrics: The new paradigm?, Chem. Mater. 22 (2010) 648–659. https://doi.org/10.1021/cm902195j.

[6]     F.P. Zhang, Q.M. Lu, J.X. Zhang, Synthesis and high temperature thermoelectric properties of $BaxAgyCa_{3−x−y}Co_4O_9$ compounds, J. Alloys Compd. 484 (2009) 550–554. https://doi.org/10.1016/j.jallcom.2009.04.144.

[7]     M. Mikami, R. Funahashi, M. Yoshimura, Y. Mori, T. Sasaki, High-temperature thermoelectric properties of single-crystal Ca3Co2O6, J. Appl. Phys. 94 (2003) 6579–6582. https://doi.org/10.1063/1.1622115.

[8]     J. Androulakis, P. Migiakis, J. Giapintzakis, La0.95Sr0.05CoO3: An efficient room-temperature thermoelectric oxide, Appl. Phys. Lett. 84 (2004) 1099–1101. https://doi.org/10.1063/1.1647686.

[9]     Y. Wang, Y. Sui, P. Ren, L. Wang, X. Wang, W. Su, H.J. Fan, Correlation between the Structural Distortions and Thermoelectric Characteristics in La 1-x A x CoO 3 (A = Ca and Sr), Inorg. Chem. 49 (2010) 3216–3223. https://doi.org/10.1021/ic902072v.

[10]    P.M. Chaikin, G. Beni, Thermopower in the correlated hopping regime, Phys. Rev. B. 13 (1976) 647–651. https://doi.org/10.1103/PhysRevB.13.647.

[11]    W. Koshibae, K. Tsutsui, S. Maekawa, Thermopower in cobalt oxides, Phys. Rev. B. 62 (2000) 6869–6872. https://doi.org/10.1103/PhysRevB.62.6869.

[12]    T. Rosenthal, M.N. Schneider, C. Stiewe, M. Döblinger, O. Oeckler, Real structure and thermoelectric properties of GeTe-Rich germanium antimony tellurides, Chem. Mater. 23 (2011) 4349–4356. https://doi.org/10.1021/cm201717z.

[13]    O. Toulemonde, N. N'Guyen, F. Studer, A. Traverse, Spin State Transition in LaCoO3 with Temperature or Strontium Doping as Seen by XAS, J. Solid State Chem. 158 (2001) 208–217. https://doi.org/10.1006/jssc.2001.9094.

[14]    D.P. Kozlenko, N.O. Golosova, Z. Jirák, L.S. Dubrovinsky, B.N. Savenko, M.G. Tucker, Y. Le Godec, V.P. Glazkov, Temperature- and pressure-driven spin-state transitions in LaCoO3, Phys. Rev. B. 75 (2007) 064422. https://doi.org/10.1103/PhysRevB.75.064422.

[15] R.X. Smith, M.J.R. Hoch, W.G. Moulton, P.L. Kuhns, A.P. Reyes, G.S. Boebinger, H. Zheng, J.F. Mitchell, Evolution of the spin-state transition with doping in La 1−x SrxCoOy, Phys. Rev. B. 86 (2012) 054428. https://doi.org/10.1103/PhysRevB.86.054428.

[16] K. Mydeen, P. Mandal, D. Prabhakaran, C.Q. Jin, Pressure- and temperature-induced spin-state transition in single-crystalline La1-x Srx CoO3 (x=0.10 and 0.33), Phys. Rev. B - Condens. Matter Mater. Phys. 80 (2009) 1–6. https://doi.org/10.1103/PhysRevB.80.014421.

[17] A. Kumar, D. Sivaprahsam, A.D. Thakur, Improvement of thermoelectric properties of lanthanum cobaltate by Sr and Mn co-substitution, J. Alloys Compd. 735 (2018) 1787–1791. https://doi.org/10.1016/j.jallcom.2017.11.334.

[18] A. Kumar, K. Kumari, B. Jayachandran, D. Sivaprahasam, A.D. Thakur, Thermoelectric properties of (1-x)LaCoO3.xLa0.7Sr0.3MnO3 composite, J. Alloys Compd. 749 (2018) 1092–1097. https://doi.org/10.1016/j.jallcom.2018.03.347.

[19] A. Kumar, M. Battabyal, A. Chauhan, G. Suresh, R. Gopalan, N. V Ravi kumar, D.K. Satapathy, Charge transport mechanism and thermoelectric behavior in Te:(PEDOT:PSS) polymer composites, Mater. Res. Express. 6 (2019) 115302. https://doi.org/10.1088/2053-1591/ab43a7.

[20] N.M.L.N.P. Closset, R.H.E. van Doorn, H. Kruidhof, J. Boeijsma, About the crystal structure of La 1− x Sr x CoO 3−δ (0≤ x ≤0.6), Powder Diffr. 11 (1996) 31–34. https://doi.org/10.1017/S0885715600008873.

[21] F. Delorme, P. Diaz-Chao, E. Guilmeau, F. Giovannelli, Thermoelectric properties of Ca3Co4O9–Co3O4 composites, Ceram. Int. 41 (2015) 10038–10043. https://doi.org/10.1016/j.ceramint.2015.04.091.

[22] D. Bhoi, N. Khan, A. Midya, M. Nandi, A. Hassen, P. Choudhury, P. Mandal, Formation of Nanosize Griffiths-like Clusters in Solid Solution of Ferromagnetic Manganite and Cobaltite, J. Phys. Chem. C. 117 (2013) 16658–16664. https://doi.org/10.1021/jp402726f.

[23] M.A. Bousnina, R. Dujardin, L. Perriere, F. Giovannelli, G. Guegan, F. Delorme, Synthesis, sintering, and thermoelectric properties of the solid solution La1–xSr x CoO3±δ (0 ≤ x ≤ 1), J. Adv. Ceram. 7 (2018) 160–168. https://doi.org/10.1007/s40145-018-0267-3.

[24] Y. Lu, Y. Li, R. Peng, H. Su, Z. Tao, D. Chen, Effect of V Substitution on the Electrical Properties of La0.5Sr0.5Co1−xVxO3 (x = 0.00–0.10) Ceramics, J. Electron. Mater. 48 (2019) 7177–7183. https://doi.org/10.1007/s11664-019-07529-4.

[25] N.V. Kosova, E.T. Devyatkina, V.V. Kaichev, Mixed layered Ni–Mn–Co hydroxides: Crystal structure, electronic state of ions, and thermal decomposition, J. Power Sources. 174 (2007) 735–740. https://doi.org/10.1016/j.jpowsour.2007.06.109.

[26] G. Tang, F. Xu, D. Zhang, Z. Wang, Improving the spin entropy by suppressing Co4+ concentration in thermoelectric Ca3Co4O9+, Ceram. Int. 39 (2013) 1341–1344. https://doi.org/10.1016/j.ceramint.2012.07.071.

[27] A. Kumar, K. Kumari, S.J. Ray, A.D. Thakur, Graphene mediated resistive switching and thermoelectric behavior in lanthanum cobaltate, J. Appl. Phys. 127 (2020) 235103. https://doi.org/10.1063/5.0009666.

[28] S. Khan, R.J. Oldman, F. Cor?, C.R.A. Catlow, S.A. French, S.A. Axon, A computational modelling study of oxygen vacancies at LaCoO3 perovskite surfaces, Phys. Chem. Chem. Phys. 8 (2006) 5207. https://doi.org/10.1039/b602753a.

[29] A. Kim, K. Song, Y. Kim, J. Moon, All Solution-Processed, Fully Transparent Resistive Memory Devices, ACS Appl. Mater. Interfaces. 3 (2011) 4525–4530. https://doi.org/10.1021/am201215e.

[30] S. Hébert, D. Flahaut, C. Martin, S. Lemonnier, J. Noudem, C. Goupil, A. Maignan, J. Hejtmanek, Thermoelectric properties of perovskites: Sign change of the Seebeck coefficient and high temperature properties, Prog. Solid State Chem. 35 (2007) 457–467. https://doi.org/10.1016/j.progsolidstchem.2007.01.027.

[31] R. Robert, L. Bocher, M. Trottmann, A. Reller, A. Weidenkaff, Synthesis and high-temperature thermoelectric properties of Ni and Ti substituted $LaCoO_3$, J. Solid State Chem. 179 (2006) 3893–3899. https://doi.org/10.1016/j.jssc.2006.08.022.

[32] K. Asai, A. Yoneda, O. Yokokura, J.M. Tranquada, G. Shirane, K. Kohn, Two Spin-State Transitions in $LaCoO_3$, J. Phys. Soc. Japan. 67 (1998) 290–296. https://doi.org/10.1143/JPSJ.67.290.

[33] K. Sato, A. Matsuo, K. Kindo, Y. Kobayashi, K. Asai, Field Induced Spin-State Transition in $LaCoO_3$, J. Phys. Soc. Japan. 78 (2009) 093702. https://doi.org/10.1143/JPSJ.78.093702.

[34] V. Vulchev, L. Vassilev, S. Harizanova, M. Khristov, E. Zhecheva, R. Stoyanova, Improving of the Thermoelectric Efficiency of $LaCoO_3$ by Double Substitution with Nickel and Iron, J. Phys. Chem. C. 116 (2012) 13507–13515. https://doi.org/10.1021/jp3021408.

[35] K. Iwasaki, T. Ito, T. Nagasaki, Y. Arita, M. Yoshino, T. Matsui, Thermoelectric properties of polycrystalline $La_{1-x}Sr_xCoO_3$, J. Solid State Chem. 181 (2008) 3145–3150. https://doi.org/10.1016/j.jssc.2008.08.017.

[36] A. Kumar, C. V Tomy, A.D. Thakur, Magnetothermopower, magnetoresistance and magnetothermal conductivity in $La_{0.95}Sr_{0.05}Co_{1-x}Mn_xO_3$ ($0.00 \leq x \leq 1.00$), Mater. Res. Express. 5 (2018) 086110. https://doi.org/10.1088/2053-1591/aad44c.

[37] E. Güneş, M.S. Wickleder, E. Müller, M.T. Elm, P.J. Klar, Improved thermoelectric properties of nanostructured composites out of $Bi_{1-x}Sb_x$ nanoparticles and carbon phases, AIP Adv. 8 (2018) 075319. https://doi.org/10.1063/1.5034525.

[38] R. Mahendiran, A.K. Raychaudhuri, Magnetoresistance of the spin-state-transition compound $La_{1-x}$, Phys. Rev. B. 54 (1996) 16044–16052. https://doi.org/10.1103/PhysRevB.54.16044.

[39] H.S. Kim, Z.M. Gibbs, Y. Tang, H. Wang, G.J. Snyder, Characterization of Lorenz number with Seebeck coefficient measurement, APL Mater. 3 (2015) 041506. https://doi.org/10.1063/1.4908244.

[40] G. Soyez, J.A. Eastman, L.J. Thompson, G.-R. Bai, P.M. Baldo, A.W. McCormick, R.J. DiMelfi, A.A. Elmustafa, M.F. Tambwe, D.S. Stone, Grain-size-dependent thermal conductivity of nanocrystalline yttria-stabilized zirconia films grown by metal-organic chemical vapor deposition, Appl. Phys. Lett. 77 (2000) 1155–1157. https://doi.org/10.1063/1.1289803.

[41] O.J. Durá, P. Rogl, M. Falmbigl, G. Hilscher, E. Bauer, Thermoelectric and magnetic properties of nanocrystalline $La_{0.7}Sr_{0.3}CoO_3$, J. Appl. Phys. 111 (2012) 063722. https://doi.org/10.1063/1.3699038.

[42] D. Sivaprahasam, S.B. Chandrasekhar, S. Kashyap, A. Kumar, R. Gopalan, Thermal conductivity of nanostructured $Fe_{0.04}Co_{0.96}Sb_3$ skutterudite, Mater. Lett. 252 (2019) 231–234.

https://doi.org/10.1016/j.matlet.2019.05.140.

[43]  A. Kumar, A. Kosonowski, P. Wyzga, K.T. Wojciechowski, Effective thermal conductivity of SrBi4Ti4O15-La0.7Sr0.3MnO3 oxide composite: Role of particle size and interface thermal resistance, J. Eur. Ceram. Soc. 41 (2021) 451–458. https://doi.org/10.1016/j.jeurceramsoc.2020.08.069.

[44]  A. Kumar, R. Kumar, D.K. Satapathy, Bi2Se3-PVDF composite: A flexible thermoelectric system, Phys. B Condens. Matter. 593 (2020) 412275. https://doi.org/10.1016/j.physb.2020.412275.

[45]  A. Kumar, K. Kumari, B. Jayachandran, D. Sivaprahasam, A.D. Thakur, Thermoelectric Properties of $(1 − x)$LaCoO3.$(x)$La$_{0.95}$Sr$_{0.05}$CoO3 composite, Mater. Res. Express. 6 (2019) 055502. https://doi.org/10.1088/2053-1591/aade73.

[46]  M.A. Bousnina, F. Giovannelli, L. Perriere, G. Guegan, F. Delorme, Ba substitution for enhancement of the thermoelectric properties of LaCoO3 ceramics ($0 \leqslant x \leqslant 0.75$), J. Adv. Ceram. 8 (2019) 519–526. https://doi.org/10.1007/s40145-019-0333-5.

[47]  F. Li, J.F. Li, Effect of Ni substitution on electrical and thermoelectric properties of LaCoO3 ceramics, Ceram. Int. 37 (2011) 105–110. https://doi.org/10.1016/j.ceramint.2010.08.024.

[48]  Y. Song, Q. Sun, Y. Lu, X. Liu, F. Wang, Low-temperature sintering and enhanced thermoelectric properties of LaCoO3 ceramics with B2O3–CuO addition, J. Alloys Compd. 536 (2012) 150–154. https://doi.org/10.1016/j.jallcom.2012.05.001.

[49]  Muhammad Umer Farooq, Ziyuan Gao, Sajid Butt, Kewei Gao, Xiaolu Pang, Hidayat Ullah Shah, Hasnain Mehdi Jafr, Asif Mahmod, Xigui Sun, Nasir Mahmood, Enhanced Thermoelectric Transport Properties of La0.98Sr0.02CoO3-BiCuSeO Composite, J. Electr. Eng. 4 (2016). https://doi.org/10.17265/2328-2223/2016.02.002.

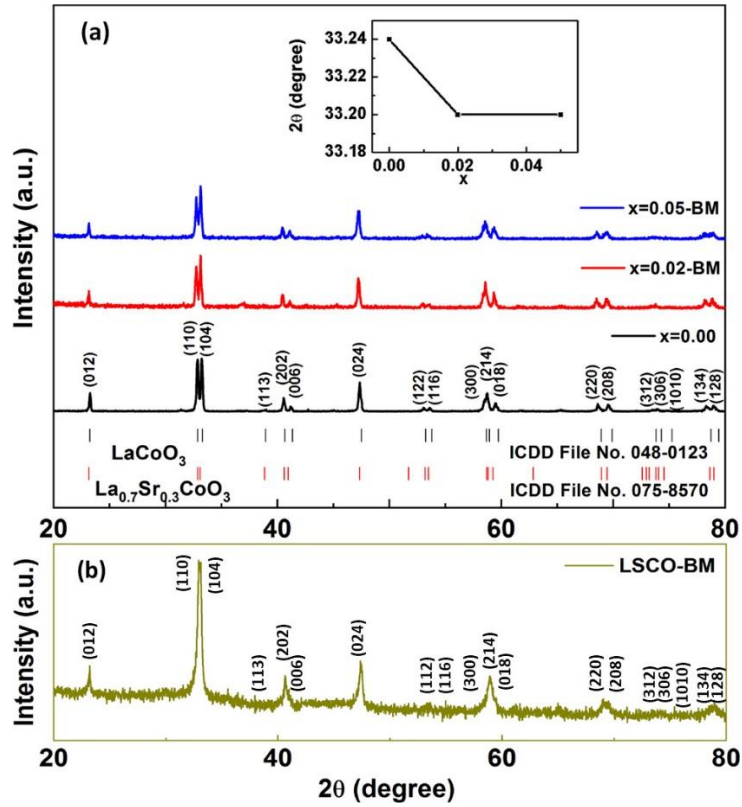

FIG. 1: (a) Powder X-ray diffraction pattern for $(1-x)$LaCoO$_3$/$(x)$ La$_{0.7}$Sr$_{0.3}$CoO$_3$ with $x$=0.00, 0.02 and 0.05 observed at room temperature. Miller indices and Bragg's position for LaCoO3 and La$_{0.7}$Sr$_{0.3}$CoO$_3$ is marked. Inset shows the change in 2θ position for $(1-x)$LaCoO$_3$/$(x)$ La$_{0.7}$Sr$_{0.3}$CoO$_3$ with $x$=0.00, 0.02-BM and 0.05-BM (b) Powder X-ray diffraction pattern for ball milled La$_{0.7}$Sr$_{0.3}$CoO$_3$.

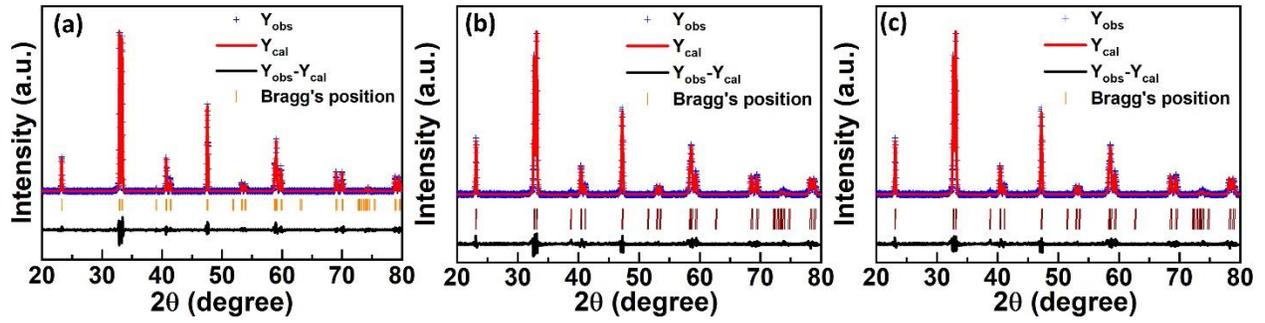

FIG. 2: Rietveld refinement pattern for $(1-x)LaCoO_3/(x)La_{0.7}Sr_{0.3}CoO_3$ (a) $x$=0.00 (b) $x$=0.02-BM (c) $x$=0.05-BM. Experimentally observed, calculated, difference curve and Bragg's position are shown.

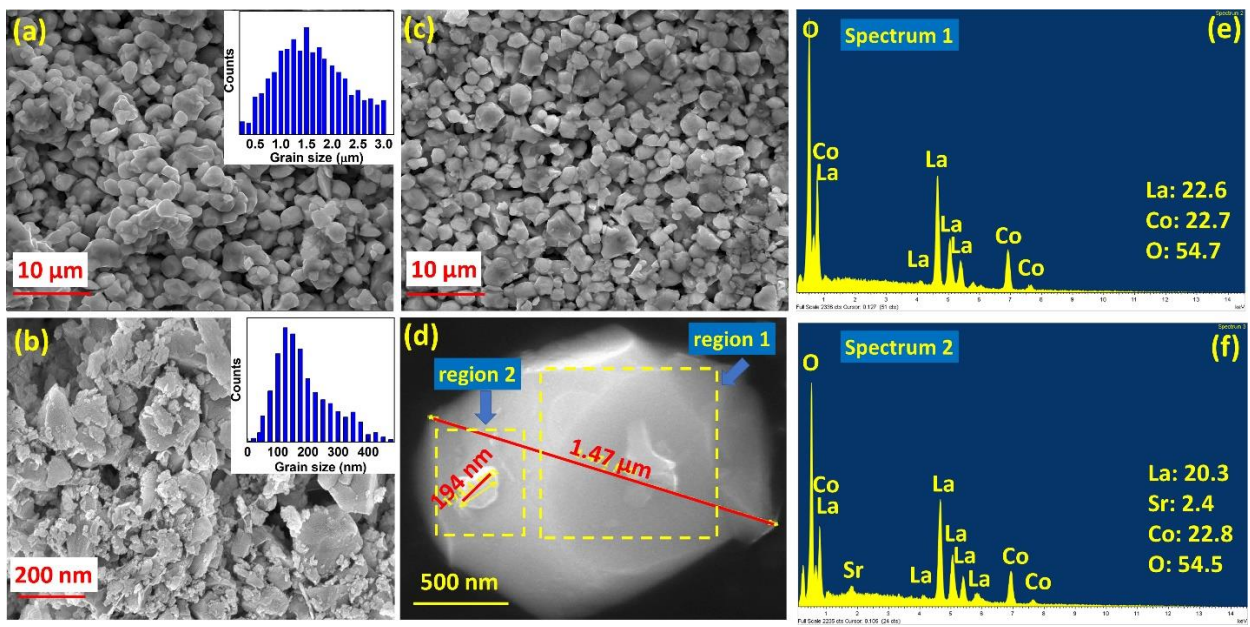

FIG. 3: FESEM image and particle size distribution (inset) for (a) LaCoO$_3$ (b) ball-milled La$_{0.7}$Sr$_{0.3}$CoO$_3$. (c) FESEM image for $(1-x)$LaCoO$_3$/$(x)$ La$_{0.7}$Sr$_{0.3}$CoO$_3$ composite powder sample with $x$=0.05-BM, (d) high magnification image for $x$=0.05-BM. Energy dispersive x-ray spectra for (e) region 1 (spectrum 1) and (f) region 2 (spectrum 2) marked in Fig.(d). The corresponding atomic % for the element present in LCO and LSCO is also shown.

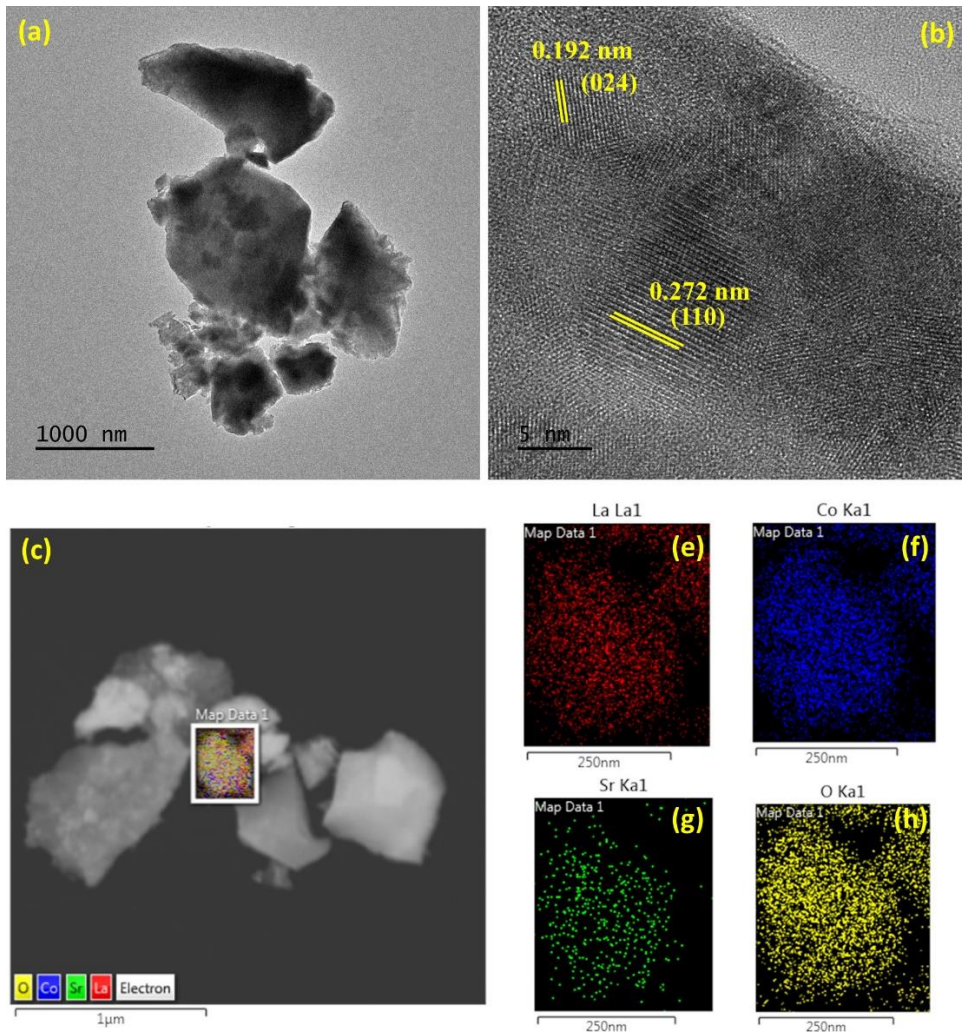

FIG. 4: Transmission electron microscope (TEM) image for (a) $(1-x)$LaCoO$_3$/$(x)$La$_{0.7}$Sr$_{0.3}$CoO$_3$ composite with $x$=0.02-BM, (b) high magnification image for $x$=0.02-BM showing the lattice spacing corresponding to LaCoO$_3$ (0.272 nm) and La$_{0.7}$Sr$_{0.3}$CoO$_3$ (0.192 nm). (c) TEM image for $(1-x)$LaCoO$_3$/$(x)$La$_{0.7}$Sr$_{0.3}$CoO$_3$ composite ($x$=0.02-BM) used for EDXS analysis, (e-h) The energy dispersive x-ray mapping for smaller particle marked in (c) for each element is shown.

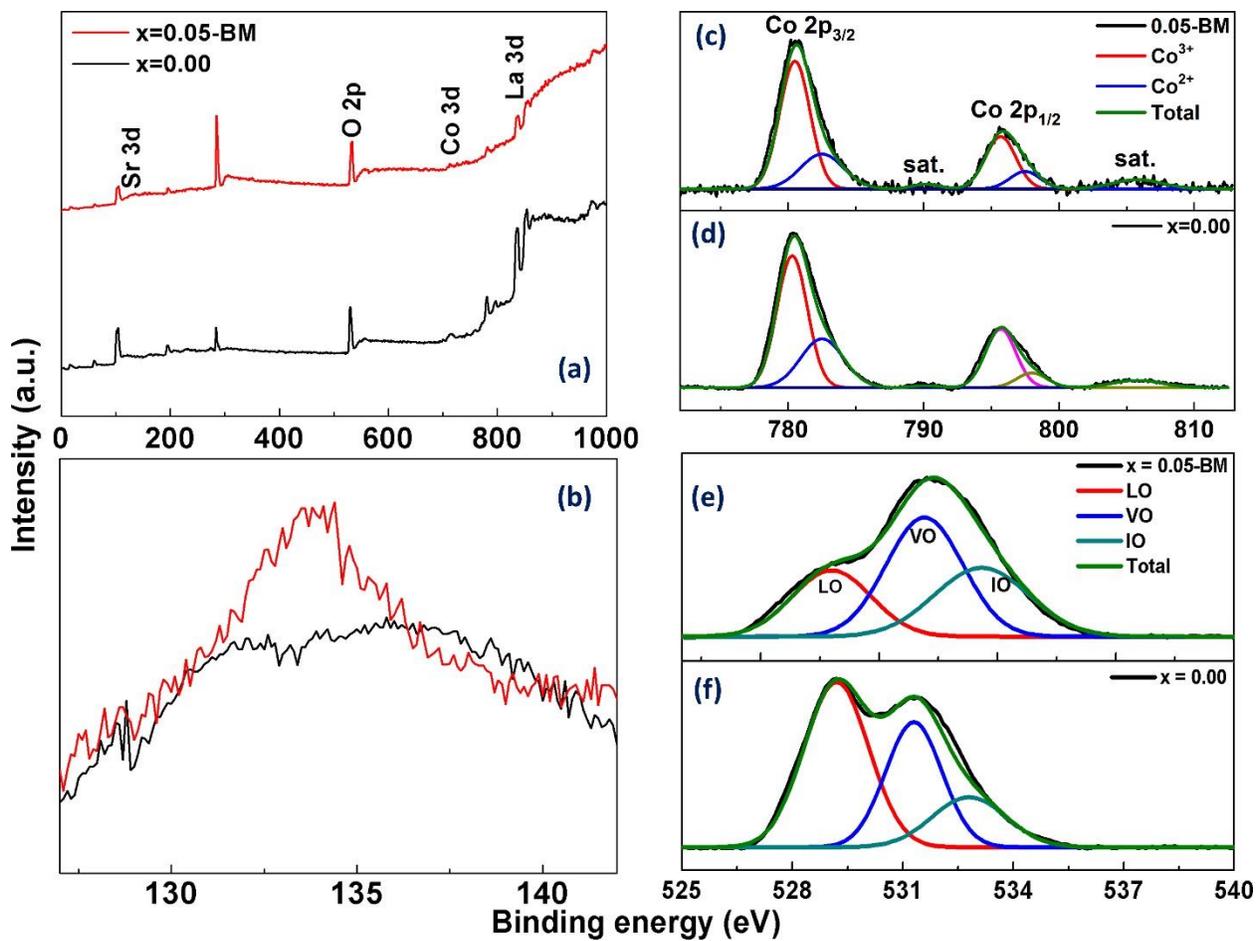

FIG. 5: XPS measurement of $(1-x)$LaCoO$_3$/$(x)$La$_{0.7}$Sr$_{0.3}$CoO$_3$ composite samples (a) full-scan (b) short-scan of Sr-3d, short-scan of Co 2p (c) $x$=0.05-BM (d) $x$=0.00, short-scan for O 1s (e) $x$=0.05-BM and (f) $x$=0.00.

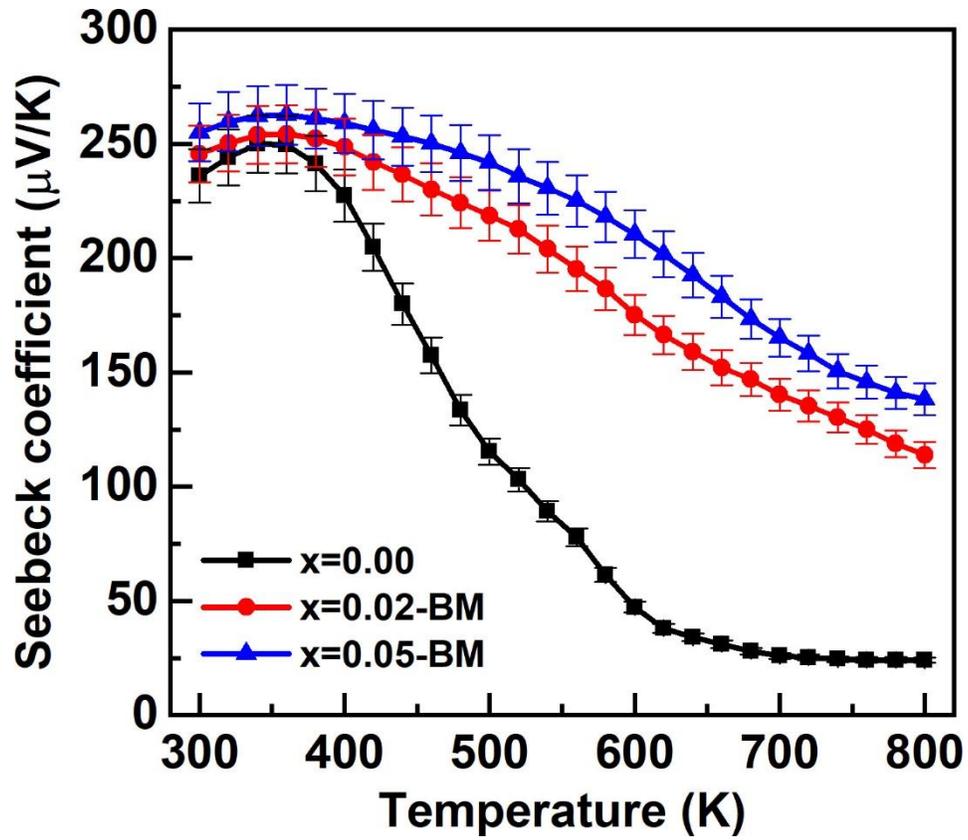

FIG. 6: Temperature variation of Seebeck coefficient (S) for $(1-x)$LaCoO$_3$/$(x)$ La$_{0.7}$Sr$_{0.3}$CoO$_3$ samples.

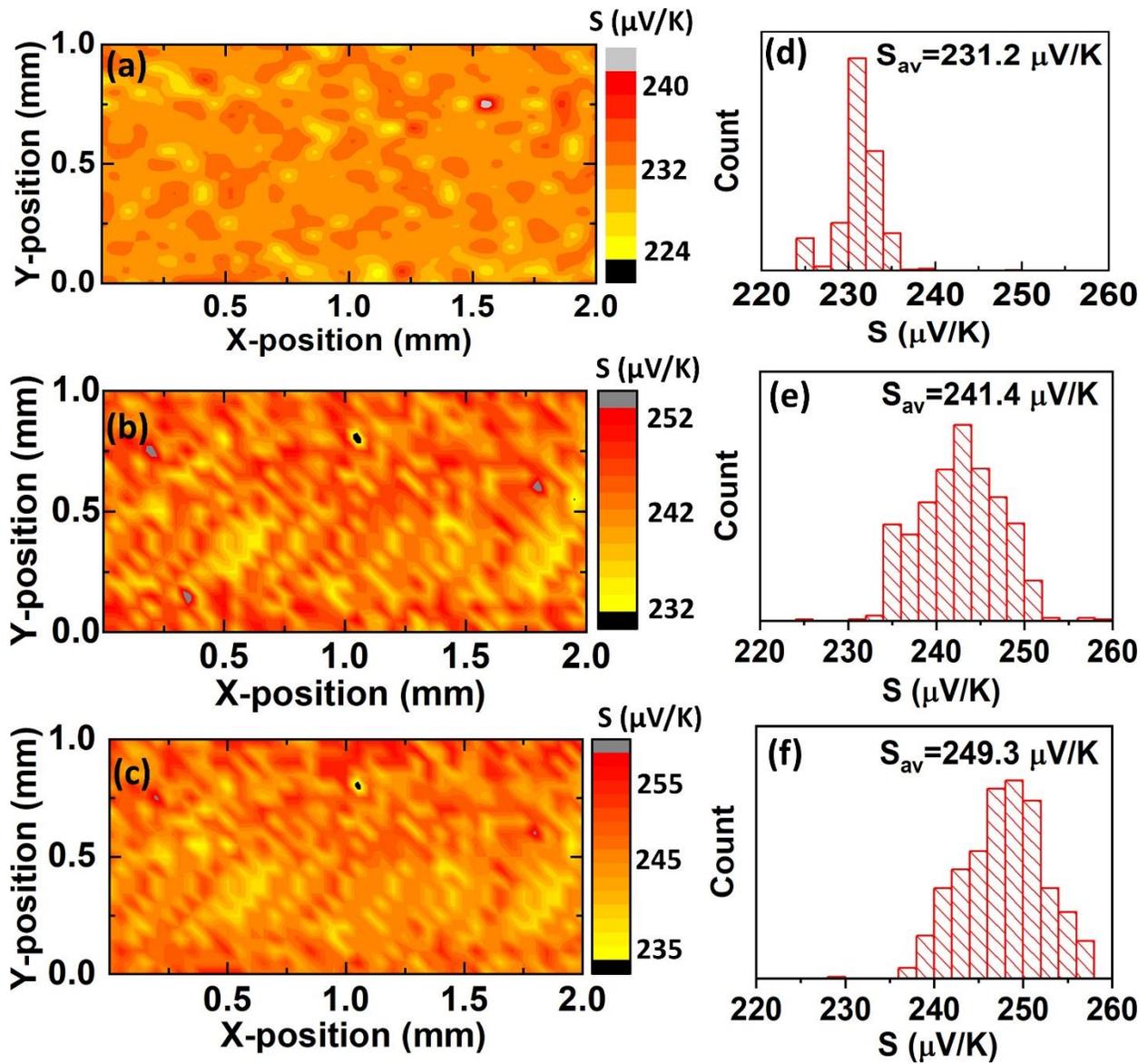

FIG. 7: Local variation of Seebeck coefficient ($S$) and corresponding histogram is shown for $(1-x)$LaCoO$_3$/$(x)$La$_{0.7}$Sr$_{0.3}$CoO$_3$ composite with (a,d) $x$=0.00 (b,e) $x$=0.02-BM (c,f) $x$=0.05- BM respectively.

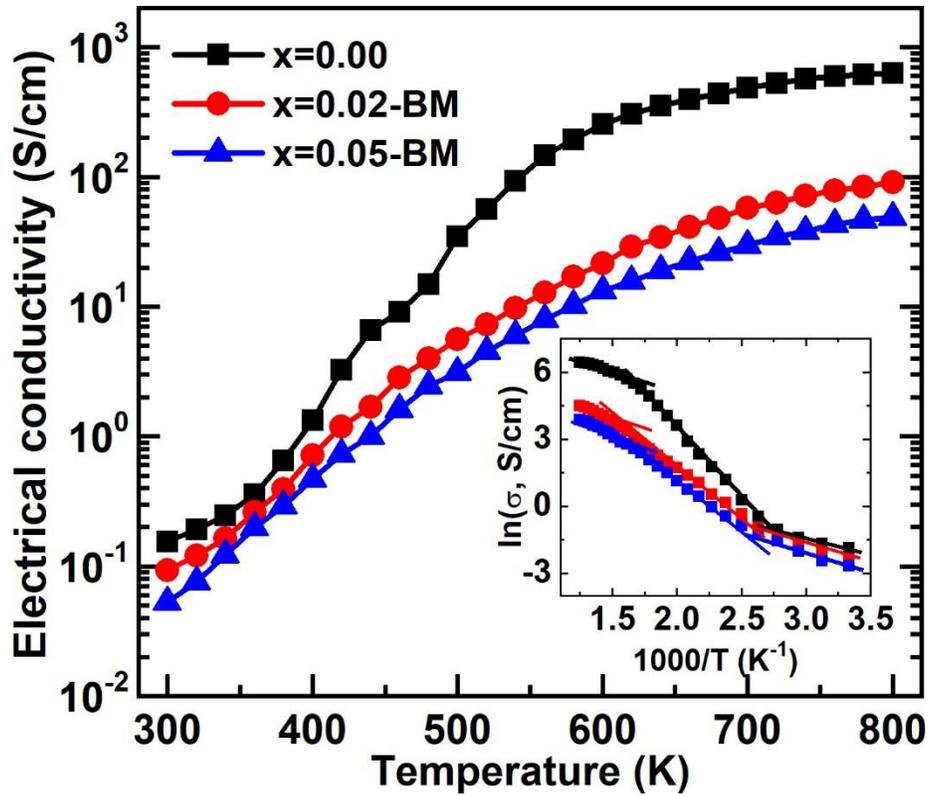

FIG. 8: Electrical conductivity ($\sigma$) as a function of temperature (T) for $(1-x)$LaCoO$_3$/$(x)$La$_{0.7}$Sr$_{0.3}$CoO$_3$ composite sample. The solid line is a guide to the eye. Inset shows the Arrhenius plot: ln $\sigma$ vs 1000/T for $(1-x)$LaCoO$_3$/$(x)$La$_{0.7}$Sr$_{0.3}$CoO$_3$ composite sample. The solid line in the inset depicts the linear fitted curve of the experimental data.

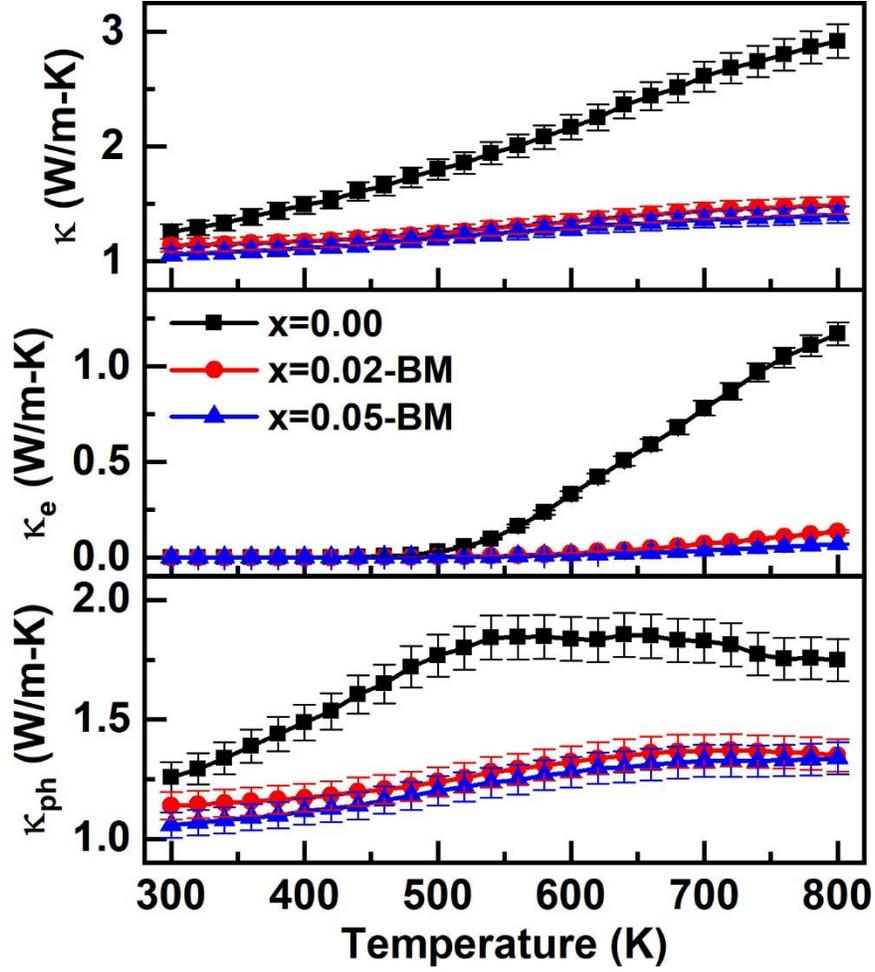

FIG. 9: Total thermal conductivity ($\kappa$), electronic thermal conductivity ($\kappa_e$) and phonon thermal conductivity ($\kappa_{ph}$) as a function of temperature (T) for $(1-x)$LaCoO$_3$/$(x)$La$_{0.7}$Sr$_{0.3}$CoO$_3$ with $x=0.00$, 0.02-BM and 0.05-BM. The Solid line is a guide to the eye.

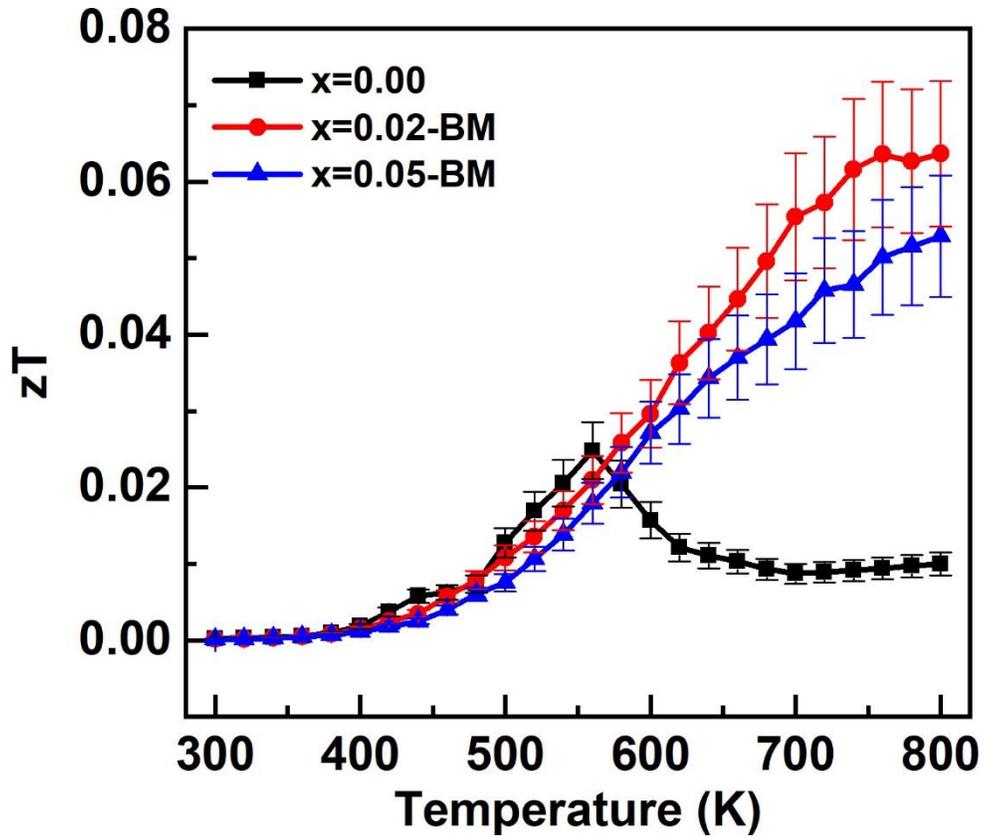

FIG. 10: Figure-of-merit zT as a function of temperature (T) for $(1-x)$LaCoO$_3$/$(x)$La$_{0.7}$Sr$_{0.3}$CoO$_3$ with $x$=0.00, 0.02-BM and 0.05-BM. The Solid line is a guide to the eye.

TABLE I: Goodness of fit ($\chi^2$), lattice parameters (a, c), and phase fraction for (1-$x$)LaCoO$_3$/($x$)La$_{0.7}$Sr$_{0.3}$CoO$_3$ (0.00 ≤ $x$ ≤ 0.05) composite sample obtained from the Rietveld refinement of the XRD patterns.

| $x$ |  | a (Å) | c (Å) | Phase % | $\chi^2$ |
|---|---|---|---|---|---|
| 0.00 | LCO | 5.439 | 13.084 | 100 | 1.64 |
| 0.02-BM | LCO | 5.439 | 13.084 | 98.4 | 1.65 |
|  | LSCO | 5.396 | 13.295 | 1.6 |  |
| 0.05-BM | LCO | 5.439 | 13.084 | 96.2 | 1.55 |
|  | LSCO | 5.396 | 13.295 | 3.8 |  |

TABLE II: Elemental compositions (atomic percentage) obtained from the energy dispersive x-ray spectroscopy (EDXS) measurement for (1-$x$)LaCoO$_3$/($x$)La$_{0.7}$Sr$_{0.3}$CoO$_3$ (0.00 ≤ $x$ ≤ 0.05) composite and La$_{0.7}$Sr$_{0.3}$CoO$_3$ sample.

| $x$ | La | Sr | Co | O |
|---|---|---|---|---|
| 0.00 | 20.88 | 0.00 | 21.12 | 58.00 |
| 0.02-BM | 21.10 | 0.15 | 21.96 | 56.79 |
| 0.05-BM | 21.03 | 1.50 | 22.05 | 55.42 |
| 1.00 | 15.90 | 5.67 | 20.96 | 57.47 |

TABLE III: Theoretical, Experimental and Relative density for (1-$x$)LaCoO$_3$/($x$)La$_{0.7}$Sr$_{0.3}$CoO$_3$ (0.00 ≤ $x$ ≤ 0.05) composite.

| $x$ | Theoretical density (g/cm$^3$) | Experimental density (g/cm$^3$) | Relative density (%) |
|---|---|---|---|
| 0.00 | 7.250 | 6.300 | 86.9 |
| 0.02-BM | 7.220 | 6.295 | 87.2 |
| 0.05-BM | 7.205 | 6.289 | 87.3 |

TABLE IV: Comparison of total thermal conductivity ($\kappa$) and figure-of-merit ($zT$) observed in similar cobaltate systems at higher temperatures

| Sample Name | $\kappa$ (W/m-K) | zT | Temp (K) | Ref. |
|---|---|---|---|---|
| LaCoO$_3$-La$_{0.3}$Sr$_{0.3}$MnO$_3$ | 2.4 | 0.023 | 800 | [18] |
| LaCoO$_3$-La$_{0.95}$Sr$_{0.05}$CoO$_3$ | 2.6 | 0.018 | 800 | [45] |
| La$_{0.95}$Ba$_{0.05}$CoO$_3$ | 5.8 | 0.013 | 750 | [46] |
| LaCo$_{0.8}$Ni$_{0.2}$O$_3$ | 3.3 | 0.012 | 673 | [47] |
| LaCoO$_3$ - (B$_2$O$_3$ - CuO) | 1.5 | 0.045 | 773 | [48] |
| La$_{0.98}$Sr$_{0.02}$CoO$_3$-BiCuSeO | 0.8 | 0.015 | 750 | [49] |
| LaCoO$_3$-La$_{0.7}$Sr$_{0.3}$CoO$_3$ | 1.4 | 0.064 | 800 | This work |